\RequirePackage{fix-cm}
\RequirePackage{mathptmx}
\documentclass[onecolumn,12pt,epjc3]{svjour3}  
\smartqed  
\RequirePackage{graphicx}
\RequirePackage[numbers,sort&compress]{natbib}
\journalname{Eur. Phys. J. C}

\usepackage{amsmath,amssymb}
\usepackage{graphicx,subfigure}
\usepackage{hyperref}
\usepackage{braket}
\usepackage{comment}
\usepackage[normalem]{ulem}

\newcommand{\be}{\begin{equation}}
 \newcommand{\ee}{\end{equation}}
 \newcommand{\bse}{\begin{subequations}}
 \newcommand{\ese}{\end{subequations}}
\newcommand{\ba}{\begin{eqnarray}}
\newcommand{\ea}{\end{eqnarray}}

\newcommand{\rhoL}{{\tilde{\rho}}}
\newcommand{\rL}{v}

\usepackage{color}
\usepackage[normalem]{ulem}  
\begin{document}

\title{How Gubser flow ends in a holographic conformal theory
}


\author{
Avik Banerjee \thanksref{addr2,addr3}
\and
Toshali Mitra  \thanksref{addr4,addr5,addr6}
\and
Ayan Mukhopadhyay \thanksref{addr7,addr8}
\and
Alexander Soloviev
\thanksref{addr1}
}



\institute{
Crete Center for Theoretical Physics, Department of Physics, University of Crete, Heraklion, Greece. \label{addr2}
           \and
           Laboratoire de Physique de l’École Normale Supérieure, ENS, Université PSL, CNRS, Sorbonne
Université, Université de Paris, F-75005 Paris, France \label{addr3}
\and
Asia Pacific Center for Theoretical Physics, Postech, Pohang 37673, Korea.\label{addr4}
\and
The Institute of Mathematical Sciences, Chennai 600113, India. \label{addr5}
\and
Homi Bhabha National Institute, Training School Complex, Anushakti Nagar, Mumbai 400094, \label{addr6}
India
\and
Instituto de F\'{\i}sica,
Pontificia Universidad Cat\'{o}lica de Valpara\'{\i}so,
Avenida Universidad 330, Valpara\'{\i}so, Chile.\label{addr7}
\and
Center for Strings, Gravitation and Cosmology, Indian Institute of Technology Madras, Chennai 600036, India \label{addr8}
\and
Faculty of Mathematics and Physics, University of Ljubljana, Jadranska ulica 19, SI-1000 Ljubljana,
Slovenia \label{addr1}
}


\maketitle

\begin{abstract}
Gubser flow is an axis-symmetric and boost-invariant evolution in a relativistic quantum field theory which is best studied by mapping $\mathbf{R}^{3,1}$ to $dS_{3}\times \mathbf{R}$ when the field theory has conformal symmetry. We show that at late de-Sitter time, which corresponds to large proper time and central region of the future wedge within $\mathbf{R}^{3,1}$, the holographic conformal field theory plasma can reach a state in which $\varepsilon = P_T = - P_L$, with $\varepsilon$, $P_T$ and $P_L$ being the energy density, transverse and longitudinal pressures, respectively. We further determine the full sub-leading behaviour of the energy-momentum tensor at late time. Restricting to flows in which the energy density decays at large transverse distance from the central axis in $\mathbf{R}^{3,1}$, we show that this decay should be faster than any power law. Furthermore, in this case the energy density also vanishes in $\mathbf{R}^{3,1}$ faster than any power as we go back to early proper time. Hydrodynamic behavior can appear in intermediate time.

\end{abstract}
\section{Introduction}
Gubser flow \cite{Gubser:2010ze,Gubser:2010ui} is a time-dependent evolution of a relativistic many-body system which is boost invariant and has a rotational symmetry about an axis. The original context of study of the Gubser flow is the evolution of QCD matter produced in central heavy ion collisions, in which case the axis of rotational symmetry is the beam axis. 

Gubser flow in conformal field theory can be studied by mapping Minkowski space ($\mathbf{R}^{3,1}$) to a product of three dimensional de-Sitter space and  the real line ($dS_{3}\times \mathbf{R}$), which makes the symmetries manifest \cite{Gubser:2010ui}. The most remarkable feature of Gubser flow is that at late de-Sitter time, $\rho$, the evolution cannot be described by relativistic hydrodynamics \cite{Gubser:2010ui,Denicol:2014xca,Denicol:2014tha,Behtash:2017wqg,Martinez:2017ibh,Denicol:2018pak,Chattopadhyay:2018fzy,Behtash:2019qtk,Dash:2020zqx}. This large $\rho$ regime is large proper time and the central region of the flow in the future wedge of the collision in Minkowski space. 

Our study of the holographic Gubser flow sets a primary example in which one can \textit{exactly} determine how a system can evolve \textit{out of} hydrodynamic regime due to expansion of some directions of space in a quantum field theory, and is relevant to understanding evolution of quantum matter in Kasner geometries \cite{Berger:2002st} describing spacetime singularities, etc. 


In kinetic theories without particle production {in the Gubser flow}, it has been shown \cite{Denicol:2014xca,Denicol:2014tha,Behtash:2017wqg,Martinez:2017ibh,Denicol:2018pak,Chattopadhyay:2018fzy,Behtash:2019qtk,Dash:2020zqx} that the large $\rho$ behavior is free-streaming where inter-particle interactions disappear. However, such ultra dilute regime is better studied in quantum field theory. Here, we show that in a holographic conformal theory, the deconfined plasma can reach a phase in which the energy density $\varepsilon$, the transverse pressure $P_T$ and the longitudinal pressure $P_L$ satisfy $P_L = - P_T$, and $\varepsilon = P_T$ in the large $\rho$ limit independently of the details of the state of the system at an early time. Remarkably, this phase is similar to the color glass condensate regime described by perturbative physics of saturated gluons \cite{Epelbaum:2013ekf,Muller:2019bwd,Berges:2020fwq}. 

We also establish a systematic late-time expansion which gives all sub-leading corrections to the colour glass condensate like behavior at late time. As in the fluid-gravity correspondence \cite{Rangamani:2009xk}, this late time behavior is deduced from demanding that the future horizon is regular at each order in the expansion. However, this late-time expansion is not a gradient expansion which breaks down at late time. It captures all systematic corrections in powers of $\exp(-\rho/L)$, where $\rho$ is the de-Sitter time and $L$ is the radius of the $S^2$ factor at $\rho = 0$. The dimensionful parameter analogous to $L$ in $R^{3,1}$ is $q^{-1}$ which is a scale set by the initial conditions, and which gives a systematic expansion in negative powers of $q\tau$ (with $\tau$ being the proper time) in $R^{3,1}$.

In physically realizable Gubser flows in $R^{3,1}$, the energy density should decay at large transverse distance from the central axis at any proper time. We show that in such flows the asymptotic late de-Sitter time behavior obtained from bulk regularity implies that the decay of the energy density at large transverse distance from the central axis in $R^{3,1}$ faster than any power. Furthermore, we also show that in the limit of early de-Sitter time $\rho \rightarrow 0$ and early proper time $\tau \rightarrow 0$, the energy density should vanish faster than any power of $\exp(\rho/L)$ and $\tau$, respectively. The systematic late de-Sitter time expansion itself carries the full information of the state (represented by the initial conditions of the gravitational solution at a finite de-Sitter time) when the symmetries of the Gubser flow are fully preserved.

We postpone discussions about the relevance of our results to heavy-ion collisions in the concluding section.

The plan of this paper is as follows. In Section \ref{Sec:Prelim}, we discuss the map from the future wedge of the collision in Minkowski space to $dS_{3}\times \mathbf{R}$, and review the breakdown of the gradient (hydrodynamic) expansion at late time. In Section \ref{Sec:HolSetup}, we doscuss the basic setup of how the Gubser flow can be studied via dual gravitational dynamics in strongly coupled large $N$ holographic conformal theories. In Section \ref{Sec:LateTime}, we demonstrate how the late de-Sitter time behavior of the Gubser flow can be determined systematically and present general perturbative solution in a perturbative expansion. In Section \ref{Sec:Beyond}, we show that by assuming that the energy profile decays in transverse directions to the axis in Minkowski space, we can also determine the early de-Sitter time behavior and how the late de-Sitter time expansion itself determines the energy density in the full future wedge of the collision in Minkowski space. We show that the energy density should decay at large transverse directions to the axis in Minkowski space faster than any power, and also vanish faster than any power as we approach the proper time $\tau =0$ which is the moment of the collision. In Section \ref{Sec:Disc}, we discuss different implications of our results and present an outlook. 

\section{Preliminaries}\label{Sec:Prelim}

For  a conformal system on $\mathbf{R}^{3,1}$, the full symmetry of the Gubser flow is $SO(3)_q\otimes SO(1,1)\otimes Z_2$. These symmetries can be made manifest by performing diffeomorphism and Weyl rescaling of $\mathbf{R}^{3,1}$ to $dS_{3}\times \mathbf{R}$ \cite{Gubser:2010ui}. The $SO(1,1)$ boost symmetry acts additively and $Z_2$, the reflection symmetry about the collision plane, acts reflexively on the $\mathbf{R}$ factor, which is physically the rapidity. Furthermore, $dS_{3}$ is a contracting and then expanding $S^{2}$ on which $SO(3)$ has a natural action.  The variable $q$ parametrizes inequivalent ways of embedding $SO(3)\otimes SO(1,1)$ in the full conformal group $SO(4,2)$. Also, $q^{-1}$ has the dimension of length and is essentially the size of the colliding systems. 

The explicit map from $\mathbf{R}^{3,1}$ to $dS_{3}\times \mathbf{R}$ is via the Milne coordinates, which are {the proper time} $\tau = \sqrt{t^2- z^2}$, the rapidity $\eta = {\rm arctan}(z/t)$, the radial coordinate $x_\perp$ and the angular coordinate $\phi$ of the plane transverse to the $z$-axis (the beam axis). In Milne coordinates, the metric on $\mathbf{R}^{3,1}$ is
\begin{equation}\label{Eq:dsM}
    {\rm d}s^2 = - {\rm d}\tau^2 + \tau^2 {\rm d}\eta^2+ {\rm d}x_\perp^2+x_\perp^2 {\rm d}\phi^2.
\end{equation}
The $dS_{3}\times \mathbf{R}$ coordinates are $\rho$, $\theta$, $\phi$ and $\eta$, where 
 \begin{eqnarray}
      \rho &=& - L \,\,{\rm arcsinh}\left(\frac{1 - q^2\tau^2 + q^2 x_\perp^2}{2q\tau}\right), \label{Eq:rho}\\ \theta &=& {\rm arctan}\left(\frac{2qx_\perp}{1 + q^2\tau^2 - q^2 x_\perp^2}\right).\label{Eq:theta}
 \end{eqnarray}
Crucially, the \textit{future wedge}, which is the causal future of the plane of the collision at $\tau =0$, is contained in $dS_{3}\times \mathbf{R}$. The corresponding metric 
\begin{align}\label{Eq:dSds2}
    {\rm d}\hat{s}^2  &= \frac{L^2}{\tau^2}{\rm d}s^2\\ &= -{\rm d}\rho^2 + L^2 \cosh^2\left(\frac{\rho}{L}\right) \left({\rm d}\theta^2 + \sin^2\theta{\rm d}\phi^2\right) +L^2 {\rm d}\eta^2,\nonumber
\end{align}
is Weyl equivalent to the flat metric. From our discussion above, it is clear that $\rho$ is invariant under $SO(3)_q\otimes SO(1,1)\otimes Z_2$, and therefore physical quantities like the energy density should depend only on $\rho$ explicitly up to the Weyl rescaling. We will denote physical quantities measured in $dS_{3}\times \mathbf{R}$ with a hat.

The conformal Ward identites for the energy-momentum tensor are:
\begin{equation}
    \hat{\nabla}_\mu \hat{T}^\mu_{\,\,\nu} = 0, \quad \hat{T}^\mu_{\,\,\mu} = \mathcal{A},
\end{equation}
where $\mathcal{A}$ is the Weyl anomaly. The $SO(3)_q\otimes SO(1,1)\otimes Z_2$ symmetries and these Ward identities imply that $\hat{T}^\mu_\nu$ should be of the form
\begin{equation}\label{Eq:TmndS}
    \hat{T}^\mu_{\,\,\nu} = \hat{t}^\mu_{\,\,\nu} +\hat{\mathcal{A}}^\mu_{\,\,\nu}
\end{equation}
where
\begin{align}
    & \hat{t}^\rho_{\,\,\rho} = -\hat{\varepsilon}\left(\frac{\rho}{L}\right), \label{Eq:e}\\
    & \hat{t}^\theta_{\,\,\theta} = \hat{t}^\phi_{\,\,\phi} = \hat{P}_T(\rho) =  -\frac{1}{2} \coth \left(\frac{\rho}{L}\right) \hat{\varepsilon} '\left(\frac{\rho}{L}\right)-\hat{\varepsilon} \left(\frac{\rho}{L}\right),\label{Eq:PT}\\
    &\hat{t}^\eta_{\,\,\eta }= \hat{P}_L(\rho) = \coth \left(\frac{\rho}{L}\right) \hat{\varepsilon}'\left(\frac{\rho}{L}\right)+3 \hat{\varepsilon}\left(\frac{\rho}{L}\right)\label{Eq:PL}.
\end{align}
Here, $'$ denotes a derivative w.r.t.~to the argument. It is easy to check that $\hat{t}^\mu_{\,\,\nu}$ is separately conserved (i.e. $\hat{\nabla}_\mu \hat{t}^\mu_{\,\,\nu} = 0$) and is traceless. The anomalous term $\hat{\mathcal{A}}^\mu_{\,\,\nu}$ (which is state-independent) is explicitly
\begin{align}\label{Eq:A}
& \hat{\mathcal{A}}^\mu_{\,\,\nu} = \frac{N^2}{32\pi^2}  {\rm diag}\left( 1,  1 ,   1 ,   -3\right)L^{-4}
\end{align}
for $\mathcal{N}=4$ $SU(N)$ super Yang-Mills-theory (SYM) \cite{Henningson:1998gx,Balasubramanian:1999re}. We note that $\mathcal{A}=\hat{\mathcal{A}}^\mu_{\,\,\mu}=0$.

It is clear from the above that the only input from the microscopic dynamics that is needed to determine the energy-momentum tensor is the evolution of the energy density $\hat{\varepsilon}(\rho)$, because the latter determines the transverse pressure $\hat{P}_T$ and the longitudinal pressure $\hat{P}_L$ via \eqref{Eq:PT} and \eqref{Eq:PL}, respectively. Weyl transformation yields the energy-momentum tensor in Minkowski space:
\begin{equation}\label{Eq:EMinK}
    {T}^\mu_{\,\,\nu} =\frac{L^4}{\tau^4} \hat{t}^\mu_{\,\,\nu}, \,\, {\rm i.e.}\,\,\varepsilon=L^4 \tau^{-4}\hat\varepsilon.
\end{equation}
Note that the anomalous term disappears.

In the hydrodynamic regime, the general form of the energy-momentum tensor given by \eqref{Eq:e}-\eqref{Eq:PL} implies that the fluid is static in the ${\rm d}S_3\times \mathbf{R}$ frame, i.e. the velocity field is of the form $\hat{u}^\rho = 1$ with other components vanishing. The hydrodynamic equations determine $\hat\varepsilon (\rho)$. For a perfect conformal fluid, $ \hat\varepsilon(\rho)=\hat\varepsilon_0 \cosh^{-8/3}(\rho/L)$, whereas for a viscous flow (see \ref{Sec:AppA} for an explicit solution), $\hat\varepsilon$ goes to a constant at large $\rho$ indicating the breakdown of the derivative expansion. However, at $\rho\sim 0$, the Knudsen number is small, and the derivative expansion is equivalent to
\begin{equation}\label{Eq:hydroexp}
    \hat\varepsilon(\rho) =\hat\varepsilon_0 - \frac{4}{3}\hat\varepsilon_0 \frac{\rho^2}{L^2}+ \frac{16}{27} H_0 \hat\varepsilon_0^{3/4} \frac{\rho^3}{L^4}+ \mathcal{O}(\rho^4),
\end{equation}
with $H_0$ being the dimensionless (and constant) shear viscosity (see \ref{Sec:AppA}). This hydrodynamic evolution is only transitory.

\section{Holographic Gubser flow}\label{Sec:HolSetup}
Any state in the universal sector of a holographic conformal gauge theory, such as $\mathcal{N}=4$ {SYM} with $SU(N)$ gauge group, has a dual description as a \textit{solution of pure classical Einstein's gravity with a negative cosmological constant in one higher spacetime dimension} in infinite 't Hooft coupling and large $N$ limits \cite{Maldacena:1997re,Witten:1998qj,Gubser:1998bc}. Such a solution should be regular, i.e. without any naked singularity, and with boundary metric $g^{(b)}_{\mu\nu}$ defined in terms of the bulk metric $G_{\mu\nu}$ via
\begin{equation}
    \lim_{r\rightarrow 0} \frac{r^2}{l^2}G_{\mu\nu} \equiv g^{(b)}_{\mu\nu},
\end{equation}
coinciding with the physical background metric on which the gauge theory lives. Above, $r$ is the holographic radial coordinate, which has an interpretation in terms of the energy scale of the dual theory \cite{Freedman:1999gp,Heemskerk:2010hk,Bredberg:2010ky,Kuperstein:2011fn,Kuperstein:2013hqa,Behr:2015aat,Behr:2015yna,Mukhopadhyay:2016fre}, while $\mu$ and $\nu$ indices represent the directions spanning the boundary of spacetime at $r=0$. The bulk cosmological constant is $\Lambda = - 6/l^2$.

Any five-dimensional gravitational solution dual to a Gubser flow of the dual gauge theory in $dS_3 \times \mathbf{R}$ can be written in the form
\begin{align}\label{Eq:ABC-metric}
    &{\rm d}s^2 = -\frac{2 l^2}{r^2} {\rm d} r {\rm d} \rho -\frac{l^2}{r^2} \left( 1 -\frac{r^2}{L^2}  + A\left(\frac{r}{L},\frac{\rho}{L}\right)\right) {\rm d}\rho^2 \nonumber\\ &+\frac{l^2(L \cosh(\rho/L) + r \sinh (\rho/L))^2}{r^2} \nonumber\\& \times e^{B\left(\frac{r}{L},\frac{\rho}{L}\right)}\,\Big( {\rm d} \theta^2 + \sin^2 \theta {\rm d} \phi^2 \Big)\nonumber\\&+ \frac{l^2}{r^2}  e^{C\left(\frac{r}{L},\frac{\rho}{L}\right)-2B\left(\frac{r}{L},\frac{\rho}{L}\right)}\,L^2{\rm d}\eta^2.
\end{align}
We have chosen the ingoing Eddington-Finkelstein gauge in which the regularity of the future horizon can be readily examined. For the boundary metric to be the metric \eqref{Eq:dSds2} on ${\rm d}S_3\times \mathbf{R}$, we need
\begin{equation}\label{Eq:BC}
    A( r=0, \rho/L) = B(r=0, \rho/L)=  C(r=0, \rho/L)=0.
\end{equation}

{The solution dual to the vacuum state corresponds to $ A = B = C=0.$ This solution is locally $AdS_5$ (it can be explicitly checked that it is the maximally symmetric metric). Although the metric dual to the vacuum of the theory in $dS_3 \times \mathbf{R}$ has all the $AdS_5$ isometries, we can readily see from \eqref{Eq:ABC-metric} that the discrete symmetry $\rho \rightarrow - \rho$ of $dS_3 \times \mathbf{R}$ at the boundary is not  preserved due to the $\sinh(\rho/L)$ term in the $\rho\rho$-component of the metric. As a result, the vacuum state itself has a monotonically growing entropy as discussed in \ref{Sec:AppD}. We note that the map from $R^{3,1}$ to $dS_3 \times \mathbf{R}$ is also defined only for the future wedge $\tau >0$, so it also prefers an arrow of time. Therefore from the Minkowski point of view, the monotonically growing entropy in the vacuum of $dS_3 \times \mathbf{R}$ could capture something similar to the entropy of the Rindler observer. We have presented some discussions on this issue in \ref{Sec:AppD}.}

Generally, as detailed in \ref{Sec:AppB}, $A$, $B$ and $C$ have radial expansions of the form 
\begin{align}\label{Eq:ABC-rad}
    & A = \sum_{n=0}^\infty a_{(n)}(\rho/L) \frac{r^n}{L^n},\,\, B = \sum_{n=0}^\infty b_{(n)}(\rho/L) \frac{r^n}{L^n},\nonumber \\ &C = \sum_{n=0}^\infty c_{(n)}(\rho/L) \frac{r^n}{L^n}
\end{align}
Plugging these in Einstein's equation, we find that the entire solution is determined just by two inputs: $a_{(1)}(\rho/L)$, which is related to a proper residual diffeomorphism that does not affect the boundary data, and $a_{(4)}(\rho/L)$ which is physical and should be chosen such that no naked singularity is present. 

Physically $a_{(4)}(\rho/L)$ determines with the energy-momentum tensor of the dual gauge theory state. The latter can be obtained systematically via the standard holographic dictionary (see \ref{Sec:AppB}) \cite{Henningson:1998gx,Balasubramanian:1999re,deHaro:2000vlm,Skenderis:2002wp}. Noting that $l^3/G_N =(2/\pi)N^2$ \cite{Maldacena:1997re}, we find that the dual energy-momentum tensor has the same form as that given by \eqref{Eq:TmndS}-\eqref{Eq:A} (including the anomalous term) with the identification
\begin{equation}\label{EQ:Ea4}
    \hat{\varepsilon}\left(\frac{\rho}{L}\right) =-\frac{3l^3}{16\pi G_N} a_{(4)}\left(\frac{\rho}{L}\right) = -\frac{3N^2}{8\pi^2 } a_{(4)}\left(\frac{\rho}{L}\right).
\end{equation}

Since a Weyl transformation at the boundary can be lifted to a bulk diffeomorphism \cite{Henningson:1998gx,deHaro:2000vlm}, we can readily obtain the Minkowski boundary metric \eqref{Eq:dsM} and the energy-momentum tensor \eqref{Eq:EMinK} via an appropriate bulk diffeomorphism.

\section{Finding the late time solution in ${\rm d}S_3\times \mathbf{R}$}\label{Sec:LateTime}

The problem of finding generic late time behavior amounts to setting up a suitable late time expansion and determining the set of conditions which lead to a regular future horizon in this expansion perturbatively. Furthermore, the solution has to be normalizable, i.e. satisfy \eqref{Eq:BC}. Due to the exponential expansion of $S^2$, the energy density in ${\rm d}S_3\times \mathbf{R}$ should dilute and we should eventually reach the vacuum. Therefore, in the dual gravitational solution \eqref{Eq:ABC-metric}, $A$, $B$ and $C$ must eventually vanish. The latter solution with $A=B=C=0$ has a horizon at $r = L$ where $\partial_\rho$ has vanishing norm. (Note this is not a Killing horizon.) Therefore at large $\rho$, the horizon should be at $r=L$ at leading order in the late time expansion, and just like in the case of fluid-gravity correspondence \cite{Rangamani:2009xk}, we should require that at each order in the perturbative expansion, the behavior of $A$, $B$ and $C$ are smooth at $r =L$ implying regularity of the future horizon.

It is instructive to first discuss the example of a massless scalar field $\Phi$ which satisfies the Klein-Gordon equation $\nabla^2 \Phi =0$, where $\nabla^2$ is the Laplacian operator in the five-dimensional metric \eqref{Eq:ABC-metric} dual to the vacuum and with $A=B=C=0$. Demanding $SO(3)_q\otimes SO(1,1) \otimes Z_2$ symmetry amounts to requiring that $\Phi$ depends only on $r$ and $\rho$. It is easy to see that the background five dimensional metric is a rational function of $\sigma = \exp(\rho/L)$. Therefore, we will expect that at late time $\Phi(r,\rho)$ should behave as $\sigma^{-\alpha} \sim  \exp(-\alpha \rho /L)$.
The consistent ansatz which fits this late time behavior (at large $\sigma$) is
\begin{equation}\label{Eq:AnsSc}
    \Phi(r,\rho) = \sigma^{-\alpha}\sum_{n=0}^\infty \phi_n(v)\sigma^{-2n}
\end{equation}
where $v = r/L$. However, if the solution is analytic at the horizon $r=L$, we should expect the behavior near $r =L$ (i.e. $v = 1$) to be given by
\begin{equation}\label{Eq:AnsHor}
    \phi_n(v) = \sum_{m=0}^\infty \lambda_{n,m} (1-v)^m,
\end{equation}
where $\lambda_{n,m}$ are pure numbers. Substituting \eqref{Eq:AnsSc} and \eqref{Eq:AnsHor} in the Klein-Gordon equation, we can readily solve all other $\lambda_{n,m}$ in terms of $\lambda_{0,m}$ which can be chosen freely. Finally, we would require the condition of normalizability, i.e.
\begin{equation}
    \phi_n(v=0) =0.
\end{equation}
Solving $\lambda_{n,0}$ explicitly in terms of $\lambda_{0,0}$, we would get
\begin{equation}
    \phi_0(v=0) =\lambda_{0,0} f_0(\alpha).
\end{equation}
The allowed values of $\alpha$ are simply the roots of $f_0(\alpha)$. Solving for $\lambda_{n,0}$ to very high orders, we get stable roots, which are $4 + 2k$, for $k = 0,1, 2, \ldots$ From the ansatz \eqref{Eq:AnsSc} and the allowed values of $\alpha$, it follows that if $\phi_{0}$ is normalizable then so is $\phi_n$ for $n\geq 1$ {provided one of the two integration constants at each order is chosen so that $\phi_n(0) =0$.} Since $\phi_0$ determines the source term in the linear equation for $\phi_1$ (and so on) for which a normalizable particular solution exists, and {the homogeneous normalizable solution also exists} since it coincides with \textit{one of the allowed values of $\alpha$, namely} $\alpha = 4 + 2k$ for $k\geq 1$, it follows that the generic late time behavior is simply \eqref{Eq:AnsSc} with $\alpha =4$. At each order in the expansion we thus have one independent integration constant after ensuring $\phi_n(0)=0$ giving a normalizable solution which is regular at the future horizon $v = 1$. Explicitly,
\begin{align}\label{Eq:phi0-phi1}
    &\phi_0(v) = \hat{o}_0 \frac{v^4}{(1+v)^4} ,\nonumber\\
    &\phi_1(v) = \hat{o}_1 \frac{v^4(3+v^2)}{3(1+v)^6}+\hat{o}_0 \frac{4v^6}{3(1+v)^6}  , \,\, {\rm etc.}
\end{align}

The expectation value (e.v.) of the marginal operator $\hat{O}$ in the gauge theory dual to the massless scalar field is essentially the $r^4$ coefficient in the radial expansion of $\Phi$. Ignoring backreaction of the dual bulk scalar field, the generic behavior of the vacuum expectation value (v.e.v.) of this operator with Gubser flow symmetries is thus
\begin{equation}\label{Eq:Ogen}
    \hat{O}(\rho) \sim \sigma^{-4} \sum_{k =0}^\infty \hat{o}_k\sigma^{-2k},
\end{equation}
where $\hat{o}_k$ are arbitrary real numbers. 

The solutions for $\phi_n(v)$ turn out to be rational functions. Furthermore, \textit{with specific choices of {integration constants} for the subleading $\phi_n$}, we can sum over the entire late time expansion as well. For instance, when the leading behaviour is $\sigma^{-4}$ (i.e. $\hat{o}_4 \neq 0$), the full summation with a specific set of choices yields
\begin{equation}\label{Eq:Sol1}
    \Phi(r,\rho) = \Gamma_4 \frac{\sigma^{-4}v^4}{(1+v)^5}\frac{(1+v)^2\sigma^2 - 3-v^2}{1+\sigma^2 + v(\sigma^2 -1)},
\end{equation}
an exact normalizable solution of the Klein-Gordon equation with an arbitrary constant coefficient $\Gamma_4$.  When the leading behaviour is $\sigma^{-6}$ (i.e. $\hat{o}_4 = 0$), we similarly obtain
\begin{equation}\label{Eq:Sol2}
    \Phi(r,\rho) = \Gamma_6 \frac{\sigma^{-6}v^4}{(1+v)^7}\frac{(1+v)^2(3+v^2)\sigma^2 - 6-8v^2-v^4}{1+\sigma^2 + v(\sigma^2 -1)},
\end{equation}
with an arbitrary constant coefficient $\Gamma_6$, and so on. Thus for each leading $\sigma^{-(4+2k)}$ behavior with $k = 0,1, 2, \ldots$, we obtain an exact normalizable solution with arbitrary constant coefficients $\Gamma_{4+2k}$. 

We readily note that both \eqref{Eq:Sol1} and \eqref{Eq:Sol2} diverge in the infinite past $\sigma =0$ (i.e. for $\rho\rightarrow -\infty$) at the horizon $v =1$, and the corresponding $\hat{O}(\rho)$ also blows up. However, since we are setting up initial conditions at an arbitrary but finite time and looking into the future, this does not bother us, as in the case of fluid-gravity correspondence where generically we get singularities on the past horizon of the leading order static black brane geometry \cite{Rangamani:2009xk,Gupta:2008th} even though the future horizon is regular for appropriate choice of transport coefficients. {This singularity in the infinite past can be cured by appropriate initial conditions.} Thus, \eqref{Eq:Ogen} gives the generic late-time behavior of $\hat{O}(\rho)$ in the vacuum. 

We can replicate the same strategy in full non-linear pure gravity with the following ansatz consistent with the gravitational equations
\begin{equation}\label{Eq:AnsGrav}
    A(r,\rho) = \sum_{n=1}^\infty \sum_{m=0}^\infty\sigma^{- n\alpha - 2m} A_{nm}(v),
\end{equation}
and with similar expansions for $B(r,\rho)$ and $C(r,\rho)$ with coefficients $B_{nm}(v)$ and $C_{nm}(v)$, respectively. {Normalizability implies that $A_{nm}(v)$, $B_{nm}(v)$ and $C_{nm}(v)$ should vanish at $v=0$.} Compared to \eqref{Eq:AnsSc}, we get a double summation here because of the non-linearity in the gravitational equations. As detailed in \ref{Sec:AppC}, \textit{both normalizability and regularity at the late-horizon $v=1$ are obtained} when $\alpha = 0, 2, 4, 6, \ldots$. However, we need to ensure that we get non-vanishing solutions which are not pure gauge. We have been able to find such physical solutions for
\begin{equation}
    \alpha = 4 + 2k, \,\, k = 0,1,2,\ldots,
\end{equation}
exactly as in the case of the massless scalar field. This implies that we can simplify the expansion \eqref{Eq:AnsGrav} to
\begin{equation}\label{Eq:AnsGrav1}
    A(r,\rho) = \sum_{k=0}^\infty\sigma^{- 4 - 2k} a_k(v), 
\end{equation}
with similar expansions for $B(r,\rho)$ and $C(r,\rho)$ with coefficients $b_k(v)$ and $c_k(v)$ respectively. Remarkably, these are of the same forms as \eqref{Eq:AnsSc} for the massless scalar case. It follows that the late de-Sitter time behavior of energy density is
\begin{equation}\label{Eq:edSGen}
    \hat{\varepsilon}(\rho) = -\frac{3l^3}{16\pi G_N}\sigma^{-4}L^{-4} \sum_{k =0}^\infty \hat{e}_k\sigma^{-2k},
\end{equation}
with $\hat{e}_k$ being arbitrary real numbers as $\hat{o}_k$ are in \eqref{Eq:Ogen}. The normalizable solutions which are regular at the late-time horizon $v=1$ and correspond to such late-time behavior \eqref{Eq:edSGen} are given by the following explicit functions appearing in \eqref{Eq:AnsGrav1} up to $\mathcal{O}(\sigma^{-8})$:
\begin{align} 
    a_0(v) &= \frac{v^4}{(1+v)^2} \hat{e}_0, \,\, b_0(v) = -\frac{v^4}{2(1+v)^4} \hat{e}_0, \,\, \nonumber \\c_0(v) &= \frac{2 v^5}{5 (1+v)^4} \hat{e}_0,\label{Eq:a0b0c0}\\
     a_1(v) &= \frac{12 \hat{e}_0 v^6}{5 (v+1)^4}+\frac{ \hat{e}_1 \left(9 v^2+10 v+10\right) v^4}{10 (v+1)^4} , \,\, \nonumber \\
     b_1(v) &= \frac{ \hat{e}_0 \left(-8 v^2+4 v-15\right) v^4}{5 (v+1)^6}+\frac{ \hat{e}_1 \left(-7 v^2+4 v-25\right) v^4}{20 (v+1)^6}, \nonumber\\ 
      c_1(v) &= \frac{4 \hat{e}_0 \left(6 v^2-7 v+14\right) v^5}{35 (v+1)^6}+\frac{ \hat{e}_1 \left(8 v^2+7 v+35\right) v^5}{35 (v+1)^6},\label{Eq:a1b1c1}\\
     a_2(v) &= -\frac{\hat{e}_0^2 v^8}{14 (v+1)^6}+\frac{2 \hat{e}_0 \left(7 v^2-4 v-6\right) v^6}{5 (v+1)^6} 
     \nonumber \\
     &+\frac{6 \hat{e}_1 \left(13 v^2+14 v+21\right) v^6}{35 (v+1)^6}
     \nonumber \\
     &+\frac{ \hat{e}_2 \left(29 v^4+70 v^3+126 v^2+70 v+35\right) v^4}{35 (v+1)^6} , \nonumber \\ 
     b_2(v) &= \frac{5 \hat{e}_0^2 v^8}{28 (v+1)^8}\nonumber \\
     &+\frac{ \hat{e}_0 \left(-91 v^4+68 v^3-364 v^2-28 v-105\right) v^4}{35 (v+1)^8} 
     \nonumber \\
     &+\frac{\hat{e}_1 \left(-82 v^4+84 v^3-588 v^2+56 v-315\right) v^4}{70 (v+1)^8}
     \nonumber\\ 
     &+\frac{ \hat{e}_2 \left(-19 v^4+20 v^3-168 v^2+28 v-140\right) v^4}{70 (v+1)^8}, \nonumber\\ 
     c_2(v) &= \frac{\hat{e}_0^2 \left(-42 v^2-250 v-225\right) v^8}{1050 (v+1)^8}
     \nonumber \\
     &+\frac{4 \hat{e}_0 \left(17 v^4-63 v^3+78 v^2-84 v-42\right) v^5}{105 (v+1)^8} 
     \nonumber \\
     &+ \frac{2 \hat{e}_1 \left(8 v^4-2 v^3+66 v^2-21 v+28\right) v^5}{35 (v+1)^8}\nonumber\\ 
     &+\frac{2 \hat{e}_2 \left(8 v^4+15 v^3+90 v^2+42 v+84\right) v^5}{105 (v+1)^8}.\label{Eq:a2b2c2}
\end{align} 

It is obvious from the above expressions that the perturbative solution is regular at $v =1$ for arbitrary values of $\hat{e}_0$, $\hat{e}_1$, 
etc.  When $\hat{e}_0 \neq 0$, and \eqref{Eq:PT} and \eqref{Eq:PL} give $\hat{\varepsilon} = \hat{P}_T = - \hat{P}_L$ in the limit $\sigma \rightarrow \infty$, i.e. for $\rho \gg L$. {We note that $\hat{e}_n$ can be of either sign. For the demonstration of monotonic entropy growth, see \ref{Sec:AppD}.} 

As detailed further in Appendix~\ref{Sec:AppC}, we have not been able to rule out that no physical solution exists for the case $\alpha =2$. We have been able to find only a pure gauge solution at the leading order in this case. If a physical solution does exist, then it will give the leading late time behavior for generic states implying that $P_T/\varepsilon \rightarrow 0$ and $P_L/\varepsilon \rightarrow 1$ exactly like in kinetic theories.  Nevertheless, our results above shows that with fine-tuned initial conditions, we can obtain a novel behavior at late time which is not realizable in kinetic theories where the energy density and the pressures should be positive. (Note even if $\alpha =2$ corresponds to physical solutions, we can get negative $\varepsilon$ and $P_L$ of equal magnitude, which is also not realizable in kinetic theories.)
 
\section{Beyond late de-Sitter time: On the behaviour in the entire future wedge of Minkowski space}\label{Sec:Beyond}
\begin{figure}[ht]
    \centering
     \includegraphics[width=0.5 \textwidth]{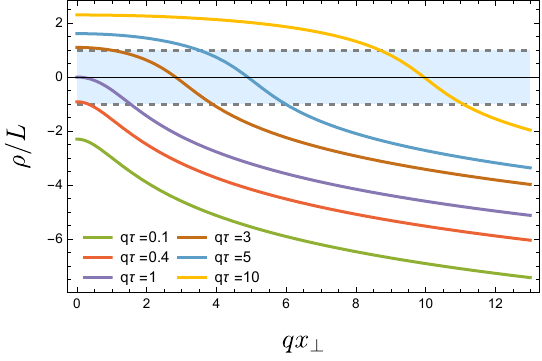}
    \caption{$\rho/L$ plotted as a function of $qx_\perp$ for various fixed values of $q\tau$. The gray dashed lines at $-1$ and 1 mark the boundary of the applicability of a hydro-like region (shaded blue).}
    \label{Fig:Story}
\end{figure}

To understand the evolution from the Minkowski point of view, we first recall how the future wedge $\tau>0$ of $\mathbf{R}^{3,1}$ maps to $dS_{3}\times \mathbf{R}$. For this purpose, it is useful to refer to Fig. \ref{Fig:Story} where $\rho/L$ is plotted as a function of $qx_\perp$ for fixed values of $q\tau$ using \eqref{Eq:rho}. We note that for $q\tau < 0.3$, the system is in the \textit{early} regime where $\rho/L$ is of large magnitude and negative. For $0.3 < q\tau < 3$, we obtain $-1<\rho/L<1$ in the central region where hydrodynamic behavior may be expected. Also, for $q\tau \gg 3$, there can be a hydrodynamic regime on any constant $\tau$ slice, within the annulus $q\tau - 1<qx_\perp<q\tau + 1$ where $-1<\rho/L<1$. The large $qx_\perp$ region is always in the early regime from the de-Sitter point of view. In the domain $q\tau \gg 3$ and $x_\perp < \tau$, we obtain $\rho/L \gg 1$, and here our late time behavior \eqref{Eq:edSGen} is valid.

It is useful to first examine the case of the free bulk scalar field for which we have exact solutions such as \eqref{Eq:Sol1}. In the latter solution, the dual operator behaves as 
\begin{equation}
    \hat{O}(\sigma) \sim \sigma^{-4}\frac{\sigma^2 -3}{\sigma^2 -1}
\end{equation}
in $dS_{3}\times \mathbf{R}$ (recall that $\sigma = \exp(\rho/L)$). The corresponding behaviour of ${O}(\sigma)$ on the future wedge of $\mathbf{R}^{3,1}$ can be obtained after the necessary Weyl scaling: $${O}(\tau,x_\perp) \sim \tau^{-4}\hat{O}(\sigma(\tau,x_\perp)).$$ Utilizing \eqref{Eq:rho}, we readily see that at large $x_\perp$, we obtain that $$O\sim \frac{x_\perp^8}{\tau^8},$$implying that the vacuum expectation value (v.e.v.) diverges at large distance from the central axis on the entire future wedge. Furthermore as $\tau \rightarrow 0$, we obtain that $$O\sim - \frac{(1+ x_\perp^2)^4}{\tau^8},$$implying that the v.e.v. diverges as we approach the initial time. The lesson is that although \eqref{Eq:Sol1} is a valid solution for the theory defined on $dS_{3}\times \mathbf{R}$, it does not render a solution which can be realized with physical initial conditions for the theory on the future wedge of Minkowski space. The same can be said about the exact solution \eqref{Eq:Sol2} as well.

Clearly, we need to constrain the behaviour in $dS_{3}\times \mathbf{R}$ so that we can map to physically realizable solutions in Minkowski space. This can be readily achieved by a re-arrangement of the late-time expansion of $\hat{O}$ given by \eqref{Eq:Ogen} in a way such that we can ensure that $O$ decays in Minkowski space at large $x_\perp$. Each term in the sum \eqref{Eq:Ogen}, which is of the form $\exp(-(4+2n)\rho/L)$ with $n$ a non-negative integer, grows with $x_\perp$ in Minkowski space as $(x_\perp^2/\tau)^{4+ 2n}$ at any fixed $\tau >0$. It is easy to see that the following expansion 
\begin{align}\label{Eq:New-O-Expansion}
   \hat{O}(\rho) = \sum_{k=0}^\infty  \left(\frac{o^a_k}{\cosh{(\rho/L)}^{4 + 2 k}} + \frac{o^b_k}{\left(2 \cosh{(\rho/L)}+\sinh{(\rho/L)} \right)^{4 + 2 k}} \right) 
\end{align}
is equivalent to the general large $\rho$ expansion \eqref{Eq:Ogen} for $\rho \gg L$, but with each term that decays along the transverse directions in Minkowski space at any fixed $\tau>0$. (Note ${\rm sech}(\rho/L)$ decays with large $x_\perp$ as $x_\perp^{-2}$ at any fixed $\tau>0$, etc.) In the second set of terms above, we have used $2\cosh(x) + \sinh(x)$ instead of $\sinh(x)$, since then the denominator does not vanish at any value of $\rho$. Also note that without the second set of terms we get an additional $\rho\rightarrow -\rho $ (de-Sitter time-reversal) symmetry which is not a feature of generic Gubser flow. Of course, there is nothing unique about the choice of basis in \eqref{Eq:New-O-Expansion}. In fact, instead of $2\cosh(x) + \sinh(x)$, we can choose $3\cosh(x) + \sinh(x)$, etc. 

We readily note that \eqref{Eq:New-O-Expansion} has double the number of coefficients when compared to the late time expansion \eqref{Eq:Ogen}, as for each $\hat{o}_k$ in \eqref{Eq:Ogen}, we have a pair $o^a_k$ and $o^b_k$ in \eqref{Eq:New-O-Expansion}. By expanding at large $\rho$, it is easy to see that
\begin{equation}\label{Eq:Late-time-coefficients}
    \hat{o}_0 = 16 \left(o^a_0 + \frac{o^b_0 }{81}\right), \quad \hat{o}_1 = 64\left(o^a_1 + \frac{o^b_1 }{729}-o^a_0 - \frac{o^b_0 }{243}\right), \,\, {\rm etc.}
\end{equation}
Each $\hat{o}_k$ is a linear combination of $o^a_k$ and $o^b_k$, and lower order terms. If we expand at early de-Sitter time $\rho\rightarrow -\infty$ (i.e. $\sigma \rightarrow 0$), we obtain that 
\begin{align}\label{Eq:Ogenearly}
   \hat{O}(\rho) = \sigma^4 \sum_{k=0}^\infty  \hat{o}^e_k \sigma^{2k},
\end{align}
with 
\begin{equation}\label{Eq:Early-time-coefficients}
    \hat{o}^e_0= 16 \left(o^a_0 + o^b_0 \right),\quad \hat{o}^e_1 = 64\left(o^a_1  +o^b_1  -o^a_0 - 3 o^b_0 \right), \,\, {\rm etc.}
\end{equation}
Each $\hat{o}^e_k$ is a linear combination of $o^a_k$ and $o^b_k$, and lower order terms. Therefore, the doubling of terms in \eqref{Eq:New-O-Expansion} is due to parametrization of both late-time and early-time behavior. Furthermore, the form \eqref{Eq:New-O-Expansion} allows for systematic expansions only when $\vert \rho \vert \rightarrow \infty$, i.e. at early and late time; and therefore this form is useful in only these limits. Our argument for considering the expansion \eqref{Eq:New-O-Expansion} shows that the early time expansion \eqref{Eq:Ogenearly} for the v.e.v. at $\sigma \sim 0$ is automatically imposed by demanding that the v.e.v. decays at large $x_\perp$ in Minkowski space and behaves as \eqref{Eq:Ogen} at late de-Sitter time. In case of the  exact solution \eqref{Eq:Sol1}, we see that the v.e.v. diverges as $\sigma^{-4}$ as $\sigma\rightarrow 0$, and therefore the v.e.v. also blows up at large $x_\perp$ in Minkowski space. If the v.e.v. has to decay in Minkowski space at large $x_\perp$ and behave as \eqref{Eq:Ogen} at late de-Sitter time, then it has to vanish as $\sigma^4$ or faster when we go back to early de-Sitter time as evident from \eqref{Eq:Ogenearly}. 

We will soon see that bulk regularity implies that $\hat{o}^e_k$ is related to $\hat{o}_k$, i.e. $o^a_k$ and $o^b_k$ are not independent of each other, and therefore the systematic late-time expansion of the v.e.v. \eqref{Eq:Ogen} in absence of sources contains the full information of initial conditions (given by the radial profile of the bulk scalar field at any fixed $\rho$). However, this inference is possible, only when in addition to bulk regularity we impose that the v.e.v. decays at large $x_\perp$ in Minkowski space so that \eqref{Eq:New-O-Expansion} is valid.
 
We readily see that in Minkowski space \eqref{Eq:New-O-Expansion} implies that at late proper time $\tau\sim\infty$ (with $x = qx_\perp$ and $t= q\tau$):
\begin{align}\label{Eq:Olatetau}
   & O (x_\perp, \tau)  = q^4 L^4\Bigg( \frac{16 (81 o^a_0 +o^b_0)}{81 t^8} 
    \nonumber\\ &+ \frac{64(729o^a_1 +o^b_1  + 729 o^a_0(-1+x^2) + 3 o^b_0 (-1 +3x^2))}{729 t^{10}}  + \mathcal{O}(t^{-12})\Bigg).
\end{align}
and at early proper time $\tau\sim 0$ (with $x = qx_\perp$ and $t= q\tau$):
\begin{align}\label{Eq:Oearlytau}
   & O (x_\perp, \tau)  = q^4 L^4\Bigg( \frac{16 (o^a_0 +o^b_0)}{(x^2+1)^4} 
    \nonumber\\ &+ \frac{64(o^a_1 +o^b_1  + o^a_0(-1+x^2)+o^b_0 (-3 +x^2))}{(x^2+1)^6} t^2 + \mathcal{O}(t^3)\Bigg).
\end{align}
The latter is similar to \eqref{Eq:Ogenearly} and we see that we get finite v.e.v. in the limit $\tau \rightarrow 0$. It also follows that from \eqref{Eq:New-O-Expansion} that we can understand the large $x_\perp$ behavior of the v.e.v. in Minkowski space via the general form (with $x = qx_\perp$ and $t= q\tau$):
\begin{align}\label{Eq:Olargex}
&O(x_\perp, \tau)  = q^4 L^4\Bigg( \frac{16 (o^a_0 +o^b_0)}{x^8} +  \frac{64 (o^a_0 +o^b_0)(-1 +t^2)}{x^{10}} \nonumber\\
& + \frac{32 (5(o^a_0 +o^b_0)+2 (- 7 o^a_0 - 9 o^b_0 + o^a_1 +o^b_1)t^2 + 5(o^a_0 +o^b_0)t^4)}{x^{12}} \nonumber\\
&+ \frac{64(-1 + t^2) (5(o^a_0 +o^b_0)+2 (- 11 o^a_0 - 17 o^b_0 +3 o^a_1 +3 o^b_1)t^2 + 5(o^a_0 +o^b_0)t^4)}{x^{14}} \nonumber\\&+ \mathcal{O}(x^{-16})\Bigg).
\end{align}
We note that both the small $\tau$ and large $x_\perp$ behavior of the v.e.v. in Minkowski space is determined systematically by \eqref{Eq:New-O-Expansion}.



In order to obtain the bulk solution of the scalar field which yields the v.e.v. which behaves as \eqref{Eq:New-O-Expansion} in de-Sitter space, we proceed with the following anstaz for the massless bulk scalar field: 
\begin{align}\label{Eq:Phi-new-expansion}
    \Phi(r, \rho) = \sum_{n=0}^\infty \frac{ \phi^a_n(v)}{\cosh{(\rho/L)}^{4 + 2 n}} + \sum_{n=0}^\infty  \frac{ \phi^b_n(v)}{\left(2 \cosh{(\rho/L)}+\sinh{(\rho/L)} \right)^{4 + 2 n}} 
\end{align}
With the above ansatz we can solve the massless Klein-Gordon equation in the background (locally) pure $AdS_5$ metric in the late time expansion ($\sigma \rightarrow\infty$) in powers of $\sigma^{-1}$ as in the previous section, and also in the early time expansion ($\sigma \rightarrow 0$) in powers of $\sigma$. At $n$-th order in these expansions, we obtained two coupled second order ordinary differential equations for $ \phi^a_n(v)$ and $ \phi^b_n(v)$. The general solutions of these equations has four integration constants. We impose that
\begin{itemize}
    \item the bulk scalar field is normalizable so that $ \phi^a_n(0) =  \phi^b_n(0) = 0$, and
    \item the solution of the bulk scalar field is regular at late time ($\sigma \rightarrow\infty$) at the future horizon $v = 1$. 
\end{itemize}
These give three conditions so that instead of four we have only one integration constant at each order which we denote as $\Gamma_n$. Remarkably, we find that imposing that the solution is regular at late time $\sigma \rightarrow\infty$ automatically implies that at each order the solution is regular at all time and even at very early time $\sigma \rightarrow 0$. We recall that this feature is absent for the exact solution \eqref{Eq:Sol1} and \eqref{Eq:Sol2}. Explicitly, the above procedure yields that 
\begin{align}\label{Eq:phi-a-b}
   &  \phi^a_0(v) = \Gamma_0 \frac{v^4}{\left(1+ v\right)^4}, \,\,\,\,\,\,\,\,\,\,   \phi^b_0(v) = -\Gamma_0 \frac{v^4}{\left(1+ v\right)^4} \nonumber \\ 
   &  \phi^a_1(v) = \frac{v^4 \left(2 \Gamma_0 \left(v^2+728 v-1085\right)+729 \Gamma_1 \left(v^2+3\right)\right)}{728 (v+1)^6},  \nonumber \\   &  \phi^b_1(v) = -\frac{3 v^4 \left(2 \Gamma_0 \left(243 v^2+728 v-119\right)+243 \Gamma_1 \left(v^2+3\right)\right)}{728 (v+1)^6}, \,\, {\rm etc.}
\end{align}

The above results can be understood in remarkably simple terms. At late time, we recover the expansion \eqref{Eq:AnsSc} with $\alpha =4$ and coefficients $\phi_n(v)$ given by \eqref{Eq:phi0-phi1} (see more below) from \eqref{Eq:Phi-new-expansion}. At early time, we obtain from \eqref{Eq:Phi-new-expansion} the expansion
\begin{equation}
    \Phi(r,\rho) = \sigma^4\sum_{n=0}^\infty \phi^e_n(v) \sigma^{2n}
\end{equation}
with
\begin{equation}
    \phi^e_0(v)= 16 \left(\phi^a_0(v) + \phi^b_0(v) \right),\quad \phi^e_1(v) = 64\left(\phi^a_1(v)  +\phi^b_1(v)  -\phi^a_0(v) - 3 \phi^b_0(v) \right), \,\, {\rm etc.}
\end{equation}
We can check that at each order $\phi^b_n(v)$ is related to $\phi^a_n(v)$ and lower order terms such that
\begin{equation}\label{Eq:phi-e-vanish}
    \phi^e_n(v) = 0, \,\, {\rm for} \,\, n= 0,1,2, \cdots, 
\end{equation}
i.e.
\begin{equation}\label{Eq:phi-e-vanish-1}
     \lim_{\sigma\rightarrow 0^+} \sigma^{-n} \Phi(r,\rho) =   \lim_{\rho\rightarrow -\infty} \exp(-n(\rho/L)) \Phi(r,\rho)   = 0, \,\, \,\, {\rm for}\,\, n = 0,1, 2, \cdots.
\end{equation}
We have checked the above to very high orders. Thus, we can conclude the late time expansion determines the entire evolution including the initial conditions when we assume that the v.e.v. decays at large transverse distance from the central axis in Minkowski space and further impose bulk regularity which necessitates the ansatz \eqref{Eq:Phi-new-expansion}. Note that at any finite value of $\rho$ (i.e. $0< \sigma <\infty$), we obtain a non-trivial profile of the bulk scalar field. 

The v.e.v. can be be readily extracted from the leading $v^4$ term of the Taylor expansion about the boundary at $v=0$ of $\phi^a_n$ and $\phi^b_n$ in \eqref{Eq:phi-a-b}, and it takes the form \eqref{Eq:New-O-Expansion} with
\begin{align}
  &  o^a_0 = \Gamma_0, \,\,\,\,\,\,\, o^b_0 = -\Gamma_0 \nonumber \\ 
  &  o^a_1 = \frac{1}{728} (2187 \Gamma_1 - 2170 \Gamma_0), \,\,\,\,\,\,\,  o^b_1 = -\frac{3 (729 \Gamma_1 -238 \Gamma_0 )}{728}, \,\, {\rm etc.}
\end{align}
Using \eqref{Eq:Late-time-coefficients}, we can extract the coefficients of the general late time expansion \eqref{Eq:Ogen} which turn out to be
\begin{align}
  \hat{o}_0= \frac{1280 \Gamma_0}{81}, \quad \hat{o}_1 = 192 \Gamma_1 -\frac{20608 \Gamma_0}{81},\,\, {\rm etc.}
\end{align}
We can similarly extract the coefficients of the general early time expansion \eqref{Eq:Ogen} using \eqref{Eq:Early-time-coefficients}. From \eqref{Eq:phi-e-vanish}, it is obvious that we should get
\begin{align}\label{Eq:Remarkable-O}
   \hat{o}^e_n = 0, \,\, \,\, {\rm for}\,\, n = 0,1, 2, \cdots.
\end{align}
i.e.
\begin{eqnarray}\label{Eq:Remarkable-O-n}
    \lim_{\sigma\rightarrow 0^+} \sigma^{-n} \hat{O}(\sigma) =   \lim_{\rho\rightarrow -\infty} \exp(-n(\rho/L)) \hat{O}(\rho)   = 0, \,\, \,\, {\rm for}\,\, n = 0,1, 2, \cdots.
\end{eqnarray}
This has further remarkable consequences. To see these, we note from \eqref{Eq:Early-time-coefficients} that this feature implies specific relations between $o^b_n$ and $o^a_n$, such as
\begin{equation}
    o^b_0 =- o^a_0, \quad o^b_1 = - o^a_1  +o^a_0 + 3 o^b_0, \,\, {\rm etc.}
\end{equation}
Substituting these in the early $\tau$ and large $x_\perp$ expansions of the v.e.v. in Minkowski space given by \eqref{Eq:Oearlytau} and \eqref{Eq:Olargex}, respectively, we find that 
\begin{eqnarray}\label{Eq:Remarkable-O-1}
    \lim_{\tau\rightarrow 0^+} \tau^{-n} O(\tau, x_\perp) =  \lim_{x_\perp\rightarrow \infty} x_\perp^{n} O(\tau, x_\perp) = 0, \,\, \,\, {\rm for}\,\, n = 0,1, 2, \cdots,
\end{eqnarray}
implying that the v.e.v. vanishes as we go back to $\tau = 0$ and at large transverse distance from the central axis faster than any positive power of $\tau$ and any negative power of $x_\perp$, respectively. 

To summarize, we solved for the bulk scalar field imposing bulk regularity and the requirement that the v.e.v. of the dual operator should decay at large transverse distance from the central axis in Minkowski space. We find that these requirements imply that the remarkable feature \eqref{Eq:Remarkable-O} that the v.e.v. and all its derivatives vanish at early de-Sitter time. This further implies that \eqref{Eq:Remarkable-O-1} must also hold. Unfortunately, we have not been able to find any explicit exact solution to all orders by choosing integration constants such that \eqref{Eq:Remarkable-O} is satisfied. So it is not possible to say anything about the radial profile of the bulk scalar field at a finite value of $\rho$ because the forms \eqref{Eq:New-O-Expansion} (and \eqref{Eq:Phi-new-expansion}) of the v.e.v. (and the bulk scalar field) give systematic expansions only at large $\vert\rho\vert$. We will come back to this issue later.

We can repeat this entire analysis for pure gravity. We assume the following form for $\hat\epsilon(\rho)$ 
\begin{align}\label{Eq:New-E-Expansion}
    \hat\varepsilon(\rho) =  \sum_{k=0}^\infty  \left(\frac{\varepsilon_k}{\cosh{(\rho/L)}^{4 + 2 k}} + \frac{\tilde{\varepsilon}_k}{\left(2 \cosh{(\rho/L)}+\sinh{(\rho/L)} \right)^{4 + 2 k}} \right),
\end{align}
which is similar to \eqref{Eq:New-O-Expansion} so that the energy density decays at large $x_\perp$ in Minkowski space and it is compatible with the late de-Sitter time expansion given by \eqref{Eq:edSGen}. This implies that the functions $A$, $B$ and $C$ in the bulk metric \eqref{Eq:ABC-metric} should take the forms
\begin{align}\label{Eq:New-ABC-Expansion}
    A(r,\rho) = \sum _{n=0} \left(\frac{a^a_n\left(v\right)}{\cosh^{2 n + 4}\left(\rho/L\right)}+\frac{a^b_n\left(v\right)}{\left(\sinh \left(\rho/L\right)+2 \cosh \left(\rho/L\right)\right)^{2 n +4}}\right), \nonumber \\
     B(r,\rho) = \sum _{n=0} \left(\frac{b^a_n\left(v\right)}{\cosh^{2 n + 4}\left(\rho/L\right)}+\frac{b^b_n\left(v\right)}{\left(\sinh \left(\rho/L\right)+2 \cosh \left(\rho/L\right)\right)^{2 n +4}}\right), \nonumber \\
     C(r,\rho) = \sum _{n=0} \left(\frac{c^a_n\left(v\right)}{\cosh^{2 n + 4}\left(\rho/L\right)}+\frac{c^b_n\left(v\right)}{\left(\sinh \left(\rho/L\right)+2 \cosh \left(\rho/L\right)\right)^{2 n +4}}\right).
\end{align}
The above form yield systematic expansions at $\rho\rightarrow\pm \infty$. In these limits, we can solve Einstein's equations systematically obtaining coupled second order ordinary differential equations at each order. Imposing normalizability and regularity at late time, we obtain unique normalizable solutions with only one integration constant at each order (after fixing the residual gauge as detailed in \ref{Sec:AppC}). Exactly, like in the case of the bulk scalar imposing bulk regularity at late time automatically implies the same at all time including $\rho\rightarrow -\infty$. Explicitly, 
\begin{align}
 & a^a_0 (v)=\frac{\gamma_0 v^4}{80 (v+1)^2}, \,\,\,\,\,\,  a^b_0 (v)= - \frac{\gamma_0 v^4}{80 (v+1)^2}, \,\,\,\,\,\,  b^a_0 = -\frac{\gamma_0 v^4}{160 (v+1)^4}, \,\,\,\,\,\,  b^b_0(v) = \frac{\gamma_0 v^4}{160 (v+1)^4}, \nonumber \\ 
 & c^a_0 (v)= \frac{\gamma_0 v^5}{200 (v+1)^4}, \,\,\,\,\,\, c^b_0 (v)= -\frac{\gamma_0 v^5}{200 (v+1)^4}, \,\nonumber\\
 & a^a_1(v) = \frac{\gamma _0 \left(2533 v^2+3650 v+10\right) v^4}{291200 (v+1)^4}+\frac{\gamma _1 \left(9 v^2+10 v+10\right)
   v^4}{7280 (v+1)^4}, \,\,\,\,\,\,  \nonumber\\&a^b_1(v) = -\frac{3 \gamma _0 \left(3271 v^2+6070 v+2430\right) v^4}{291200 (v+1)^4}-\frac{\gamma _1 \left(9 v^2+10
   v+10\right) v^4}{7280 (v+1)^4}, \,\,\,\,\,\, \nonumber\\
 & b^a_1(v) = -\frac{\gamma _0 \left(3979 v^2+7292 v-35\right) v^4}{582400 (v+1)^6}-\frac{\gamma _1 \left(7 v^2-4 v+25\right)
   v^4}{14560 (v+1)^6}, \,\,\,\,\,\, \nonumber \\
 &b^b_1(v) = \frac{9 \gamma _0 \left(1251 v^2+2428 v+805\right) v^4}{582400 (v+1)^6}+\frac{\gamma _1 \left(7 v^2-4
   v+25\right) v^4}{14560 (v+1)^6},\nonumber \\
   &c^a_1 (v)= \frac{\gamma _0 \left(4352 v^2+5131 v-2569\right) v^5}{1019200 (v+1)^6}+\frac{\gamma _1 \left(8 v^2+7
   v+35\right) v^5}{25480 (v+1)^6}, \,\,\nonumber \\
 & c^b_1(v) = -\frac{9 \gamma _0 \left(1616 v^2+2835 v+847\right) v^5}{1019200 (v+1)^6}-\frac{\gamma _1 \left(8 v^2+7
   v+35\right) v^5}{25480 (v+1)^6}, \,\,\,\,\,\, {\rm etc.}
\end{align}
which reproduces the functions in \eqref{Eq:a0b0c0}, \eqref{Eq:a1b1c1} and \eqref{Eq:a2b2c2} at late de-Sitter time. At early time, we obtain exactly like in the case of the scalar field that
\begin{equation}
    \lim_{\sigma\rightarrow 0^+}\sigma^{-n}A(r, \sigma) =\lim_{\sigma\rightarrow 0^+}\sigma^{-n}B(r, \sigma)= \lim_{\sigma\rightarrow 0^+}\sigma^{-n}C(r, \sigma)= 0, \,\, {\rm for}\,\, n = 0,1,2, \cdots.
\end{equation}

From the above, we readily find that 
\begin{align}
   &\varepsilon_0 = \gamma_0/80, \,\,\,\,\,\,\,\,\,\,\,\,\,\,\, \tilde{\varepsilon}_0 = -\gamma_0/80 \nonumber \\
   & \varepsilon_1 = \frac{\gamma_0 + 40 \gamma_1}{29120}, \,\,\,\,\,\,\,\,\,\,\,\,\,\,\, \tilde{\varepsilon}_1 = - \frac{729 \gamma_0 + 40 \gamma_1}{29120}, \,\, {\rm etc.}
\end{align}
The coefficients of the late-time expansion \eqref{Eq:edSGen} turn out to be
\begin{align}
   \hat{e}_0 = \frac{16 \gamma_0}{81},\quad \hat{e}_1 = -\frac{8 \left(363\gamma_0-40 \gamma_1\right)}{3645}, \,\, {\rm etc.}
\end{align} 
As in the case of the scalar field we obtain that
\begin{align}\label{Eq:Remarkable-e}
   \lim_{\sigma\rightarrow 0}\sigma^{-n}\hat\varepsilon(\sigma) = 0, \,\, \,\, {\rm for}\,\, n = 0,1, 2, \cdots.
\end{align}
which implies that in Minkowski space
\begin{eqnarray}\label{Eq:Remarkable-O-1}
    \lim_{\tau\rightarrow 0^+} \tau^{-n} \varepsilon(\tau, x_\perp) =  \lim_{x_{\perp} \rightarrow \infty} x_\perp^{n} \varepsilon(\tau, x_\perp) = 0, \,\, \,\, {\rm for}\,\, n = 0,1, 2, \cdots.
\end{eqnarray}
To summarize, the requirements of regularity of the bulk metric (at late time) and that the dual energy density decays at large $x_\perp$ in Minkowski space imply that the energy density should satisfy the remarkable feature that it and all its derivatives vanish at early de-Sitter time. The latter feature further implies that the energy density vanishes in Minkowski space both in the limits $\tau \rightarrow 0^+$ and $x_\perp \rightarrow \infty$ faster than any positive power of $\tau$ and any negative power of $x_\perp$, respectively. We can also conclude that when the energy density decays at large $x_\perp$ on the future wedge of Minkowski space, the late de-Sitter time expansion itself contains the full information of the initial conditions set at any finite value of $\rho$ or $\tau$ since the coefficients $\tilde{\varepsilon}_n$ in \eqref{Eq:New-E-Expansion} get related to $\varepsilon_n$, and therefore there is no independent information needed from early time to describe the energy density on the entire future wedge.

{The feature of the Gubser flow that we can deduce information about the initial conditions by constraining that the energy density (and expectation values of other operators) should decay at large transverse distance from the central axis is certainly distinctive as such an extrapolation is not possible for the Bjorken flow in which there is homogeneity in the transverse directions. Essentially, this feature is a consequence of the symmetries of the Gubser flow which ties the $x_\perp$ dependence to $\tau$ dependence as both arise via the dependence on the de-Sitter time $\rho$.}

We note that the form \eqref{Eq:New-E-Expansion} is compatible with the hydrodynamic expansion \eqref{Eq:hydroexp} at $\rho \sim 0$. However, this form of the energy density and also the form \eqref{Eq:New-ABC-Expansion} of the functions in the bulk metric admit systematic expansions only at large $\vert\rho\vert$, and therefore we cannot systematically match at $\rho = 0$ or any finite value of $\rho$. {The same can also be said about any resummation of the late-time expansion. For instance, one can readily note that for the exact solutions of the massless scalar $\Phi(v,\rho)$ given by \eqref{Eq:Sol1} and \eqref{Eq:Sol2}, the late time expansion in $\sigma^{-1}$ converges only for $\sigma^2 > (1-v)/(1+v)$ for any fixed $v$. Thus, matching with hydrodynamic evolution, which can be valid only around $\rho\sim 0$ (i.e. $\sigma\sim 1$), is difficult.} Irrespective of whether hydrodynamic evolution is realized around $\rho \sim 0$, it is difficult to determine the nature of the metric at a finite value of $\rho$ because unlike the case of the scalar field, it is apriori even difficult to estimate where the late and early time expansions converge.

Therefore, only by solving Einstein's equations numerically (say following the method of characteristics developed in \cite{Chesler:2010bi}), we would be able to determine which initial conditions set at a finite value of $\rho$ or $\tau$ would be realizable physically, i.e. those in which the energy density does decay at large $x_\perp$. We do not expect that for all such initial conditions the hydrodynamic expansion \eqref{Eq:hydroexp} would actually be realized at $\rho \sim 0$.
Numerical simulations are also necessary to establish convergence properties of the late time and early time expansions obtained from \eqref{Eq:New-E-Expansion} and contrast it with the trans-series forms discussed in \cite{Behtash:2017wqg,Denicol:2018pak,Behtash:2019qtk,Dash:2020zqx,Soloviev:2021lhs}.

\section{Discussion}\label{Sec:Disc}

We have provided some exact results on how relativistic quantum matter evolves in Gubser flow in holographic strongly coupled large $N$ conformal gauge theories. The most generic behaviour is yet to be established and further progress can be made by employing numerical relativity in this context. Numerical analyses are needed to confirm many of the extrapolations which we have made for the behavior on the entire future wedge in Minkowski space from the late de-Sitter time behavior. However, the late de-Sitter time behavior can be rigorously established just from bulk regularity alone.

The main implications of our results is that in flows with such special symmetries, a relativistic quantum system can evolve out of hydrodynamics into a new regime that is \textit{independent} of the initial conditions. Such evolution can therefore reveal many fundamental features of the underlying microscopic theory. Particularly, we show that when the symmetries of the Gubser flow are preserved, the late de-Sitter time evolution can lead to a color glass condensate-like phase which is \textit{independent} of the initial conditions. Of course, this phase is very different from the perfect fluid state which is realized for generic initial conditions which violate the symmetries of the Gubser flow.

Another implication of our results is that the initial conditions which realize such flows with special symmetries are further constrained beyond these symmetries. We have shown that in Gubser flows in holographic strongly coupled large $N$ conformal gauge theories, the energy density should vanish faster than any power of the proper time $\tau$ as we go back to $\tau =0$, the boundary of the future wedge. It is very unlikely that such a behavior can be realized by colliding gravitational shock waves (which represent collision of energy lumps in the dual gauge theory) as studied in \cite{Chesler:2010bi}. The bulk radial profiles of these shocks should be supported sufficiently far away from the boundary so that the dual energy density (given by the fourth radial derivative of the $A$ function in the bulk metric \eqref{Eq:ABC-metric}) and its proper time derivatives should vanish at the moment of collisions, and this could as well imply that the radial profiles of these shocks actually vanish. Therefore, instead of being generated by ``asymptotic states,'' the Gubser flow naturally could be smoothly glued to pure vacuum.\footnote{Since $\varepsilon(x_\perp, \tau)$ and all its $\tau$-derivatives vanish at $\tau =0$, one can argue via a Fefferman-Graham type expansion as well that the bulk metric at $\tau=0$ should be close to pure $AdS_5$. This could be another alternative way to prove that the bulk solution can be smoothly glued to pure $AdS_5$ outside the future wedge.} The non-trivial behavior in the future wedge evolving back to the vacuum arises from self-consistent pressure gradients. Of course, we need to numerically confirm the extrapolation we have made to the entire future wedge from the late de-Sitter time behavior.

It would be also interesting to investigate bulk solutions which behave as a Gubser flow in the vicinity of a specific azimuthal direction but is otherwise homogeneous in transverse directions, so that it corresponds to a Bjorken flow approximately. Such a solution can be interpreted as a jet embedded in an expanding medium and also created at early proper time with non-trivial initial conditions. The existence of such solutions could indicate that Gubser flow can be applied to collective flow arising in jets if not the flow of the full quark gluon plasma. Gubser flow indeed has recently been successfully applied for understanding two and four particle correlations in jets with large charge multiplicity arising in pp-collisions \cite{Taghavi:2019mqz,ALICE:2023lyr} (in \cite{Taghavi:2019mqz}, it has been argued that four particle cumulants can be related to fluctuations in initial conditions).

It would be interesting to understand these issues in QCD:
\begin{itemize}
    \item Is there a generic nature of the final phase of the evolution of Gubser flow in de-Sitter time which is independent of the initial conditions, and if so, then what are the characteristics of this phase? 
    \item What kind of constraints on the initial conditions lead to such flow with special symmetries like Gubser flow (beyond the necessary symmetries)?
    \item Can such initial conditions be realized perhaps with greater likelihood in jets and other substructures within the full bulk evolution of the system?
\end{itemize}
These questions are relevant for the phenomenology of heavy-ion collisions. Therefore, we plan to study Gubser flow in holographic non-conformal/confining gauge theories \cite{Gursoy:2015nza,Chattopadhyay:2021ive,Jaiswal:2021uvv,Chen:2021wwh} and in semi-holographic \cite{Iancu:2014ava,Banerjee:2017ozx,Kurkela:2018dku,Mitra:2020mei,Mitra:2022xtb} scenarios and see how both, the initial conditions and the late de-Sitter regime for the Gubser flow distinguishes confining behaviour from an emergent infrared critical point.

 The advantage of studying Gubser flow in a holographic theory is that we can understand the quantum information theoretic aspects of such a scenario in which a many-body system can escape hydrodynamization, especially the fundamental reason why and how despite lack of hydrodynamization the evolution can reach a phase independent of the initial conditions. Many novel aspects of quantum thermodynamics \cite{RevModPhys.93.035008} can be understood via explicit computations of entanglement measures (see for example \cite{Kibe:2021qjy,Banerjee:2022dgv}).  A preliminary discussion on the entropy production captured by the growth of the area of the horizons, and puzzles regarding its interpretation has been presented in \ref{Sec:AppD}. 


\begin{acknowledgements}
It is a pleasure to thank Jorge Casalderrey-Solana, Suresh Govindarajan, Sa{\v s}o Grozdanov, Matti J\"{a}rvinen, David Mateos, 
Giuseppe Policastro, Alexandre Serantes and Amitabh Virmani for discussions. We thank Sa{\v s}o Grozdanov, Edmond Iancu, Matti J\"{a}rvinen, David Müller, Mike Strickland, Victor Roy and Amitabh Virmani for comments on the manuscript. Both AB and AM acknowledge support from IFCPAR/CEFIPRA funded project no 6304-3. Research of AB is partially supported by the European MSCA grant HORIZONMSCA-2022-PF-01-01 and by the H.F.R.I call “Basic research Financing (Horizontal support of all Sciences)” under the National Recovery and Resilience Plan “Greece 2.0” funded by the European Union– NextGenerationEU (H.F.R.I. Project Number: 15384). The research of AM is also supported by the Center of Excellence initiative of the Ministry of Education of India, and the new faculty seed grant of IIT Madras. AS is supported by funding from Horizon Europe research and innovation programme under the Marie Skłodowska-Curie grant agreement No. 101103006 and the project
N1-0245 of Slovenian Research Agency (ARIS). TM has been supported by an appointment to the JRG Program at the APCTP through the Science and Technology Promotion Fund and Lottery Fund of the Korean Government. TM has also been supported by the Korean Local Governments -- Gyeong\-sang\-buk-do Province and Pohang City -- and by the National Research Foundation of Korea (NRF) funded by the Korean government (MSIT) (grant number 2021R1A2C1010834).\end{acknowledgements}

\appendix

\section{Viscous Hydrodynamics}\label{Sec:AppA}

The conformal viscous Gubser flow has been studied in \cite{Gubser:2010ui}. The conformal perfect fluid energy-momentum tensor is $ \hat{t}^{\mu\nu}_0 = \varepsilon(T) (4 u^\mu u^\nu + g^{\mu\nu})$, where $\varepsilon(T) \propto T^4$ is determined by the equation of state in terms of the temperature $T$. Including the first order viscous correction, we have
\begin{align}
\hat{t}^{\mu\nu} = \hat{t}^{\mu\nu}_0 -2 \eta(T) \sigma^{\mu\nu},
\end{align}
where $\sigma^{\mu\nu}$ is the shear stress tensor defined as
\begin{align}
    \sigma^{\mu\nu} =\frac{1}{2}\Delta^{\alpha \mu} \Delta^{\beta \nu} \left(\nabla_{\alpha} u_{\beta} + \nabla_{\beta} u_{\alpha} \right) - \frac{1}{3} \nabla_\lambda u^\lambda \Delta^{\mu\nu},
\end{align} 
and $ \eta(T)\propto T^3$ is the shear viscosity. It is useful to define $ H_0 = \eta(T)\hat{\varepsilon}^{-3/4}(T)$ which is a constant in a conformal field theory and is determined by microscopic dynamics. Above, $\Delta^{\mu\nu} = u^\mu u^\nu + g^{\mu\nu}$. The conservation of the energy-momentum tensor (i.e. the hydrodynamic equations) for the Gubser flow in ${\rm d}S_3 \times \mathbf{R}$ frame is simply 
\begin{align}
\hat{\varepsilon}'(\rhoL ) +\frac{8}{3} \tanh (\rhoL ) \hat{\varepsilon} (\rhoL )  - \frac{16}{9} H_0  \tanh ^2(\rhoL ) \hat{\varepsilon} (\rhoL )^{3/4}= 0,
\end{align}
with $\rhoL = \rho/L$. The solution of this equation is
\begin{align}
\varepsilon(\rhoL) &=  \frac{\text{sech}^4(\rhoL )}{531441}  \Big(4 H_0 L^{-1} \sinh ^3(\rhoL ) {\cosh ^{1/3}(\rhoL )}  \nonumber \\ & _2F_1\left(\frac{7}{6},\frac{3}{2};\frac{5}{2};-\sinh ^2(\rhoL )\right)+27 \hat{\varepsilon}_0^{1/4} {\cosh^{1/3} (\rhoL )}\Big){}^4.
\end{align}
The viscous energy density $\hat{\varepsilon}$ in the limit $\rhoL \rightarrow 0$ gives the derivative expansion \eqref{Eq:hydroexp} as mentioned in the main text. However, in the limit $\rhoL \rightarrow \infty$ the energy density has the following expansion dominated by viscous term at the leading order: 
\begin{align}
    \hat{\varepsilon}(\rhoL) = \frac{16}{81} H_0^4 L^{-4} + \mathcal{O}(\cosh(\rhoL)^{-2/3}) + \ldots
\end{align}
This indicates the breakdown of derivative expansion in the large $\rhoL$ limit.

\section{Holographic renormalization}\label{Sec:AppB}

For compactness, we define $\rhoL=\rho/L$ and also use the variable $v$ with $\rL=r/L$ as in the main text. We can solve for the metric functions $ A(\rL,\rhoL),  B(\rL,\rhoL) $ and $  C(\rL,\rhoL)$ in the radial expansion near the boundary $\rL = 0$ as indicated in \eqref{Eq:ABC-rad} and obtain,
\begin{align}\label{Eq:near_boundaryExpA}
 &A(\rL,\rhoL) =  a_{(1)}\left(\rhoL\right)\rL +\frac{1}{4} \left(a_{(1)}^2\left(\rhoL\right) -4 a_{(1)}'\left(\rhoL\right)\right) \rL^2 \nonumber \\ &+ a_{(4)}\left(\rhoL\right) \rL^4 + \ldots, \\
\label{Eq:near_boundaryExpB}
&B(\rL,\rhoL) = a_{(1)}(\rhoL) \cosh ^2(\rhoL) ~ \rL \nonumber \\ & -  \frac{1}{4} a_{(1)}(\rhoL) \left( a_{(1)}(\rhoL)  + 4 \tanh (\rhoL)\right) ~ \rL^2 \nonumber \\  & + \frac{1}{12} a_{(1)}(\rhoL) \Big( \left(a_{(1)}(\rhoL)  + 3 \tanh \left(\rhoL \right) \right)^2  + 3 \tanh ^2\left(\rhoL\right) \Big) ~ \rL^3
\nonumber \\   & + b_{(4)}(\rhoL) ~ \rL^4 +  \ldots ,\\
\label{Eq:near_boundaryExpS}
& C(\rL,\rhoL) = 3  a_{(1)}(\rhoL) ~ \rL - \frac{1}{4  }  a_{(1)}(\rhoL)\Big(8 \tanh (\rhoL) + 3 a_{(1)}(\rhoL) \Big) ~ \rL^2 \nonumber \\ & + \frac{1}{4} a_{(1)}(\rhoL) \left( \left(a_{(1)}(\rhoL) + 2 \tanh (\rhoL) \right)^2 + 4 \tanh ^2(\rhoL) \right) ~ \rL^3 \nonumber \\ & + s_{(4)}(\rhoL) ~ \rL^4 + \ldots ,
\end{align}
where $ a_{(1)}\left(\rhoL\right)$ is an arbitrary function. The functions $  b_{(4)}\left(\rhoL\right)$ and $  s_{(4)}\left(\rhoL\right)$ are determined by $ a_{(4)}\left(\rhoL\right)$ and $a_{(1)}\left(\rhoL\right)$ by the constraints of Einstein's equations:
\begin{align}\label{Eq:constraint}
& s_{(4)}\left(\rhoL\right) = -\frac{1}{32} a_{(1)}\left(\rhoL\right)  \Big(  a_{(1)}^2\left(\rhoL\right) + 4 \tanh ^2\left(\rhoL\right) \Big) \nonumber  \\ & \Big( 3  a_{(1)}\left(\rhoL\right) + 16 \tanh \left(\rhoL\right) \Big) + \frac{9}{8} a_{(1)}^2\left(\rhoL\right)  \tanh^2 \left(\rhoL\right),
\nonumber \\
& a_{(4)}'\left(\rhoL\right) =  \frac{8}{3} \tanh \left(\rhoL\right) \Big(b_{(4)}\left(\rhoL\right)-a_{(4)}\left(\rhoL\right)\Big) \nonumber 
\\ & + \frac{1 }{12} \Big(a_{(1)}^3\left(\rhoL\right) + 4 a_{(1)}\left(\rhoL\right)\tanh^2 \left(\frac{\rho }{L}\right) \Big) \nonumber \\ & \Big( a_{(1)}\left(\rhoL\right) + 8 \tanh \left(\rhoL\right) \Big) + \frac{5}{6} a_{(1)}^2\left(\rhoL\right) \tanh^2\left(\rhoL\right).
\end{align}
We find that the radial expansions and thus $A$, $B$ and $C$ are entirely determined by the two functions $ a_{(4)}\left(\rhoL\right)$ and $a_{(1)}\left(\rhoL\right)$.

However, $a_{(1)}\left(\rhoL\right)$ is a residual gauge freedom and can be set to zero by the diffeomorphism $v\rightarrow v + f(\rhoL)$ which preserves the ingoing Eddington-Finkelstein gauge. Furthermore, the latter is a proper diffeomorphism, meaning that $a_{(1)}\left(\rhoL\right)$ does not affect boundary data. Indeed extracting the dual energy-momentum tensor $\langle T^{\mu }_{\,\nu} \rangle$ at the boundary  from the renormalized on-shell gravitational action following \cite{Henningson:1998gx,Balasubramanian:1999re,deHaro:2000vlm} and using \eqref{Eq:constraint}, we find that $a_{(1)}\left(\rhoL\right)$ disappears. Explicitly, we obtain
\begin{align}\label{Eq:boundarygubser_stress_tensor}
\langle T^{\mu }_{\,\nu} \rangle =  
\begin{pmatrix}
\hat{t}^\rho_{\,\rho }  & 0 & 0 & 0 \\ 
0 & \hat{t}^\theta_{ \,\theta} &0 &0 \\ 
0 & 0& \hat{t}^\phi _{\,\phi}  &0 \\
0& 0& 0& \hat{t}^\eta_{\,\eta}
\end{pmatrix} + \mathcal{A}^\mu_{\,\nu}
\end{align}
where 
\begin{align}
  & \hat{t}^\rho_{\,\rho} \equiv- \hat{\varepsilon}\left(\rhoL\right) = \frac{3l^3}{16 \pi G_N } a_{(4)}\left(\rhoL\right), \nonumber \\
 &  \hat{t}^\theta_{\,\theta} \equiv \hat{P_T}\left(\rhoL\right) = \frac{3l^3}{16 \pi G_N }  \left( \frac{1}{2} \coth \left(\rhoL\right) a_{(4)}'\left(\rhoL\right)+  a_{(4)}\left(\rhoL\right)\right), \nonumber \\
 &  \hat{t}^\phi_{\,\phi} \equiv \hat{P_T}\left(\rhoL\right) =  \hat{t}^\theta_{\,\theta}, \nonumber \\
 & \hat{t}^\eta_{\,\eta} \equiv  \hat{P_L}\left(\rhoL\right) =  - \frac{3l^3}{16 \pi G_N }\Big( \coth \left(\rhoL\right) a_{(4)}'\left(\rhoL\right)+ 3 a_{(4)}\left(\rhoL\right) \Big).
\end{align} 
and $ \mathcal{A}^\mu_{\,\nu}$ is the Weyl anomaly given by
\begin{eqnarray}
\hat{\mathcal{A}}^{\mu}_{\,\nu } &=& \frac{l^3}{128 \pi G_N } \Big(\frac{4}{3} R^\mu_{\,\nu} R - 2 R^{\mu}_{\,\kappa} R^\kappa_{\,\nu} - \delta^{\mu}_{\,\nu } \Big( \frac{1}{2} R^2 - R_{\kappa \sigma} R^{\kappa \sigma}\Big) \Big)  \nonumber \\
&=& \frac{l^3}{64 \pi G_N } diag\left( 1,  1 , 1,   -3 \right) L^{-4},
\end{eqnarray} 
where the curvatures refer to those of the background metric \eqref{Eq:dSds2} for the dual theory (which is also the boundary metric of the five-dimensional bulk spacetime). With the identification $l^3/G_N = 2 N^2/\pi$, we obtain \eqref{Eq:A}.

\section{Finding values of $\alpha$ for pure gravity}\label{Sec:AppC}
Here we give details of how $\alpha$ has been determined replicating the strategy for the massless scalar field discussed in the main text. The ansatz \eqref{Eq:AnsGrav} is a double expansion in $\sigma$. 
Assuming analytic behavior at the future horizon $r=L$ (i.e. $v=1$), we obtain the near horizon expansion (with numerical coefficients $a_{nm,p}$)
\begin{align}
A_{nm}(v) = \sum_{p=0}^{\infty} a_{nm,p}(1-v)^p, 
\end{align}
and similarly for $B_{nm}(v)$ and $C_{nm}(v)$ in terms of the numerical coefficients $b_{nm,p}$ and $c_{nm,p}$, respectively. Solving the gravitational equations in this near-horizon expansion, we find that $A_{10}(v)$, $B_{10}(v)$ and $C_{10}(v)$ are determined by the three integration constants $a_{10,0}$, $b_{10,0}$ and $c_{10,0}$. Explicitly,
\begin{align}\label{Eq:Anbsol}
  & A_{10}(v) =   a_{10,0} \left(1 - (\alpha+1) (1-v) + \mathcal{O}((1-v)^2)\ldots \right) \nonumber \\  & + b_{10,0} \left(-2 \alpha (1-v) + \mathcal{O}((1-v)^2) + \ldots \right) \nonumber \\ &+ c_{10,0} \left(\frac{1}{2} \alpha(1+\alpha) (1-v) +\mathcal{O}((1-v)^2) + \ldots \right), 
\end{align}
and also similarly for $B_{10}(v)$ and $C_{10}(v)$. 

Normalizable solutions can be matched to the near-boundary expansion \eqref{Eq:near_boundaryExpA}-\eqref{Eq:near_boundaryExpS}. However, the solution of $A_{10}(v)$ given by \eqref{Eq:Anbsol} when expanded near the boundary $v=0$ reads as
\begin{align}\label{Eq:A10nb}
     A_{10}(v) &= P(\alpha) (a_{10,0} + (1+\alpha) (2 b_{10,0}-c_{10,0})) \\
     &+ \mathcal{O}(v) + \ldots\nonumber
\end{align}
where $P(\alpha)$ is polynomial in $\alpha$. Therefore, for normalizability, $\alpha$ should be the roots of $P(\alpha)$ and these are $\alpha = 2k$, with $k = 0, 1, 2, \ldots$. This can be found by performing the expansion \eqref{Eq:Anbsol} to high orders and then re-expanding it as a Taylor series about $v=0$ as in the case of the massless scalar. 

For $k=0$, $A_{10}$, $B_{10}$ and $C_{10}$ vanish. For the case $k=1$ we have been able to find a solution which corresponds to the boundary expansion \eqref{Eq:near_boundaryExpA}-\eqref{Eq:near_boundaryExpS} with $a_{(1)}(\rhoL)\neq 0$ but $a_{(4)}(\rhoL)=0$. Since $a_{(1)}(\rhoL)$ corresponds to a pure residual (proper) gauge freedom, this solution for the case $k=1$ yields just a pure gauge deformation. However, we cannot rule out this case completely as another physical solution may exist. For $k = 2, 3, \ldots$, we are able to find physical solutions as reported in the main text.

To see how we reach these conclusions, it is useful to use the residual gauge freedom to set $a_{(1)}(\rhoL)=0$. In this case, the near boundary expansions \eqref{Eq:near_boundaryExpA}-\eqref{Eq:near_boundaryExpS} start from $v^4$. For the comparison with near-horizon expansions for $A_{10}$, $B_{10}$ and $C_{10}$ we should further impose a late-time expansion of $a_{4}(\rhoL)$ of the form
\begin{equation}\label{Eq:a4lte}
  a_{(4)}(\rhoL) =   \sum_{n=1}^\infty \sum_{m=0}^\infty\sigma^{- n\alpha - 2m} a_{(4)nm}
\end{equation}
that follows from \eqref{Eq:AnsGrav} with constant coefficients $a_{(4)nm}$. Setting the coefficients of $v$ and $v^2$ to zero in \eqref{Eq:A10nb}, we get two relations between the three integration constants $a_{10,0}$, $b_{10,0} $ and $c_{10,0}$. For $k=0$, these simply set $a_{10.0} =0$ and then requiring that the boundary values of $B_{10}$ and $C_{10}$ to vanish (so that we get normalizable solutions) we obtain $b_{10.0} = c_{10.0} = 0$. Then $A_{10}$, $B_{10}$ and $C_{10}$ have to vanish. For $k\geq 2$, setting the coefficients of $v$ and $v^2$ to zero in \eqref{Eq:A10nb} give two relations to determine $b_{10,0}$ and $c_{10,0}$ in terms of $a_{10,0}$. Using these values of $b_{10,0}$ and $c_{10,0}$, we find a perfect agreement with the near boundary expansions \eqref{Eq:near_boundaryExpA}-\eqref{Eq:near_boundaryExpS} corresponding to normalizable solutions where the residual gauge freedom has been fixed with $a_{(1)}=0$. The remaining integration constant $a_{10,0}$ simply determines the leading term in $a_{(4)}(\rhoL)$ in \eqref{Eq:a4lte}, i.e. $a_{(4)10}$. For the allowed values of $\alpha$ which are $4+ 2n$, with $n=0,1,2, \ldots$ the ansatz \eqref{Eq:AnsGrav} can be simplified to \eqref{Eq:AnsGrav1}, as mentioned before. (Then $a_{(4)10}$ identified with $\hat{e}_0$ up to a numerical constant, etc.) The case $k=1$ is somewhat tricky because the series expansion of the equations near the horizon themselves do not have unique solutions. One way of solving it leads to a pure gauge solution. We have not been able to show that physical solutions do not exist for this case.

At higher orders in the late de-Sitter time expansion in case of $\alpha = 4+ 2n$ and $n= 0,1,2,\ldots$, we obtain normalizable solutions because the lower orders source only normalizable particular solutions, while the homogeneous solutions are also normalizable with a new arbitrary integration constant at each order since $\alpha =6, 8, \ldots$ are allowed to be the leading order behavior also (when the lower order coefficients vanish). This argument is similar to the case of the massless bulk scalar. We thus establish that \eqref{Eq:edSGen} gives realizable late de-Sitter time expansion for the dual energy density.

\section{Examining the entropy}\label{Sec:AppD}
Let us first examine the entropy $S$ in the de-Sitter frame. The entropy $S$ of the black hole given by \eqref{Eq:ABC-metric} is given by the area $\Sigma_H$ of the apparent or event horizon located at $\rL=\rL_h(\rhoL)$
\begin{align}\label{Eq:entropy}
    S = \frac{\Sigma_H}{4 G_N}
\end{align}
where $G_N$ is the Newton constant. Here we will examine the event horizon. The area can be computed via $\Sigma_H = \int_{\rL = \rL_h} \sqrt{\gamma} ~ d \theta ~ d \phi ~ d\eta$ with $\gamma$ being the induced metric on the spatial sections of the event horizon (generated by a congruence of null geodesics). Explicitly,
\begin{align}
    \gamma = l^6\frac{e^{C(v_h(\rhoL),\rhoL )}\sin^2\theta\left(\cosh \left(\rhoL \right)+ v_h(\rhoL) \sinh \left(\rhoL \right) \right)^4}{v_h^6(\rhoL)}.
\end{align}
Here $\theta \in [0,\pi]$, $\phi \in [0,2\pi]$ and $\eta \in [-\infty, \infty]$. Due to the infinite extent of $\eta$, it is better to define the entropy per unit rapidity which is
\begin{align}
    \frac{dS}{ d \eta} = \pi \frac{l^3}{G_N} \frac{e^{\frac{C(v_h(\rhoL),\rhoL )}{2}} \left(\cosh \left(\rhoL \right)+ v_h(\rhoL) \sinh \left(\rhoL \right) \right)^2}{v_h^3(\rhoL)}.
\end{align}
The corresponding entropy density per unit rapidity reads
\begin{align}
    \frac{ds}{ d \eta} =  \frac{1}{4\pi L^3 \cosh^2\left( \rhoL\right)}\frac{dS}{ d \eta}.
\end{align}
The factor of $4\pi \cosh^2\left( \rhoL\right)$ is the area of the sphere at the boundary (measured by the background metric \eqref{Eq:dSds2}). The location of the event horizon $\rL_h(\rhoL)$ can be determined by the radial null geodesic equation, i.e.
\begin{align}\label{Eq:geodesic}
    \frac{{\rm d} v_h(\rhoL)}{{\rm d} \rhoL } + \frac{1 - v_h^2 (\rhoL)+A(v_h(\rhoL),\rhoL)}{2} = 0
\end{align}
with the condition that $v_h(\rhoL = \infty) =1$, since the horizon coincides with that of the solution dual to the vacuum in the limit $\rhoL\rightarrow\infty$. 

\begin{figure}[ht]
    \centering
    \includegraphics[width=0.4 \textwidth]{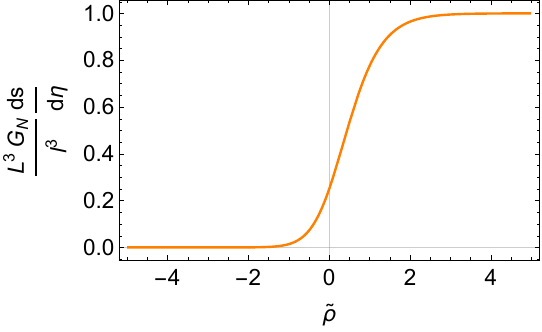}
    \caption{Vacuum entropy density per unit rapidity plotted as a function of $\rhoL$. The figure shows that at late-time $ds/d\eta$ goes to a constant.}
    \label{Fig:vacuum_entropy}
\end{figure}

In the state dual to the vacuum, we simply have
\begin{align}\label{Eq:vaceng}
    &\frac{d s}{ d \eta} = \frac{l^3}{4 L^3 \cosh^2\left(\rhoL \right) G_N}  \left(\cosh \left(\rhoL \right)+ \sinh \left(\rhoL \right) \right)^2,\nonumber\\
    &=\frac{l^3}{L^3 G_N} \times \frac{1}{4} \left(1+ \tanh \left(\rhoL \right) \right)^2.
\end{align}
This implies that the entropy density per unit rapidity monotonically increases from zero to a constant value with the de-Sitter time as shown in Fig. \ref{Fig:vacuum_entropy}.

At this point, it would seem strange to associate an entropy to the vacuum state in the Gubser flow {both from the de-Sitter and Minkowski future wedge points of view. In the latter case, this is the entropy associated with an accelerated observer like the Rindler observer. The appearance of entropy in the de-Sitter vacuum is more subtle and has been discussed recently in \cite{Chandrasekaran:2022cip}. Although the bulk metric dual to the vacuum in de-Sitter space is locally $AdS_5$, note that bulk diffeomorphism can produce a non-trivial entropy as in the case of the map from pure three-dimensional anti-de-Sitter space ($AdS_3$) to the Banados Teitelboim Zanelli (BTZ) black hole \cite{Banados:1992wn} (however the physical interpretation of the entropy in our case is more similar to that in \cite{Emparan:1999gf}.)}

The bulk dual of aGubser flow admits the late-time expansion \eqref{Eq:AnsGrav1}, and thus the location of the event horizon which is a solution of \eqref{Eq:geodesic} with the condition $v_h(\rhoL = \infty) =1$, admits the following expansion
\begin{align}
    v_h(\rhoL) = 1 + \sum_{m=0}^\infty v_{h,m} e^{- (4 + 2m)\rhoL}
\end{align}
at late time. Using the explicit perturbative solution for the metric given by \eqref{Eq:a0b0c0}-\eqref{Eq:a2b2c2}, we obtain
\begin{align}
    v_{h,0} = \frac{\hat{e}_0}{40}, \quad  v_{h,1} = \frac{24\hat{e}_0 + 29\hat{e}_1}{2240}, \quad \ldots
\end{align}
and
\begin{align}
 L^3  \frac{G_N}{l^3} \frac{ds}{d \eta} = 1 - 2 e^{- 2 \rhoL} + 3\left(1 -  \frac{\hat{e}_0}{80} \right) e^{- 4 \rhoL} + \ldots
\end{align}
 In the case of the vacuum with $0=\hat{e}_0=\ldots$, the above series has alternating signs. However, the function is monotonically increasing (and assuming the value $1$ in the limit $\rhoL\rightarrow\infty$) as evident from the exact expression \eqref{Eq:vaceng} (that is plotted in Fig. \ref{Fig:vacuum_entropy}). The monotonic growth of the entropy density per unit rapidity is not affected perturbatively and it reaches the constant vacuum value at late time. It has been shown in \cite{Maldacena:2012xp} that the entanglement entropy can scale as the volume in de-Sitter space with a numerical factor which has a maximal value.
If $\hat{e}_0$ is positive, then the energy density at leading order is negative, and $ds/d\eta$ is less than in the vacuum. On the other hand if $\hat{e}_0$ is negative, then the energy density at leading order is positive, and $ds/d\eta$ is larger than that in the vacuum. Here, it is also useful to note that it has been argued that the entropy of an excited state in de-Sitter space is expected to be less than that of the vacuum \cite{Chandrasekaran:2022cip}. It will be useful to understand the case of the holographic Gubser flow better with a clear algebraic interpretation of $ds/d\eta$ computed here.



\bibliography{gubser.bib}
\bibliographystyle{spphys}

\end{document}